\documentclass[letterletter, 10 pt, journal, twoside]{ieeetran}
\input epsf
\usepackage[square,numbers]{natbib}
\usepackage{xcolor,soul,framed} 
\colorlet{shadecolor}{yellow}
\usepackage{graphicx}
\usepackage{amsmath}
\newtheorem{theorem}{\textbf{Theorem}}
\newtheorem{lemma}{\textbf{Lemma}}
\newtheorem{example}{\textbf{Example}}

\newtheorem{remark}{\textbf{Remark}}
\newtheorem{definition}{\textbf{Definition}}

\newtheorem{proposition}{\textbf{Proposition}}
\newtheorem{assumption}{\textbf{Assumption}}
\usepackage{url}
\newenvironment{proof}{{{\bf Proof:}}}{\hfill $\square$\par}
\usepackage{multirow} 
\usepackage{xcolor}
\usepackage{multirow}
\usepackage{setspace}
\usepackage{url}
\usepackage{subfigure}
\usepackage{fancyhdr}
\usepackage{amsmath}
\usepackage{multirow}
\usepackage{amssymb }
\usepackage{color}
\usepackage{graphics} 
\usepackage{graphicx}
\usepackage{algorithm} 
\usepackage{algorithmic} 
\usepackage{subeqnarray}
\usepackage{cases}
\usepackage{geometry}
\usepackage{setspace}

\renewcommand{\baselinestretch}{0.971} \normalsize
\begin{document}

\title{{\Large Total Unimodularity and Strongly Polynomial Solvability of Constrained Minimum Input Selections for Structural Controllability: an LP-based Method}}

\author{Yuan Zhang, Yuanqing Xia, and Yufeng Zhan

  \thanks{This work was supported in part by the China Postdoctoral Innovative Talent Support Program under Grant BX20200055, and the
National Natural Science Foundation of China under Grant 62003042. The authors are with School of Automation, Beijing Institute of Technology, Beijing, China. Email: {\{zhangyuan14,xia\_yuanqing,yu-feng.zhan\}@bit.edu.cn}. Corresponding author: Y. Xia.}
 }
\maketitle 

\begin{abstract} This paper investigates several cost-sparsity induced optimal input selection problems for structured systems. Given are an autonomous system and a prescribed set of input links, where each input link has a non-negative cost. The problems include, selecting the minimum cost of input links, and selecting the input links with the smallest possible cost with a bound on their cardinality, all to ensure system structural controllability. Current studies show that in the dedicated input case (i.e., each input can actuate only a state variable), the former problem is polynomially solvable by some graph-theoretic algorithms, while the general nontrivial constrained case is largely unexploited. We show these problems can be formulated as equivalent integer linear programming (ILP) problems. Subject to a certain condition on the prescribed input configurations that contains the dedicated input one as a special case, we demonstrate that the constraint matrices of these ILPs are {\emph{totally unimodular}}. This property allows us to solve those ILPs efficiently simply via their linear programming (LP) relaxations, leading to a unifying algebraic method for these problems with polynomial time complexity. It is further shown those problems could be solved in strongly polynomial time, independent of the size of the costs and cardinality bounds. Finally, an example is provided to illustrate the power of the proposed method.

\end{abstract}
\IEEEoverridecommandlockouts

%
\IEEEpeerreviewmaketitle

\section{Introduction} \label{intro-sec}

Input/output (I/O) selection, or actuator/sensor placement for a control system to possess certain performances, is of great importance for control design. Most of the I/O selection problems are challenging due to their combinatorial nature \cite{V.D.2001A}. Over the past decades, significant achievements have been achieved in understanding the structure and computational complexity of various I/O selection problems concerning a wide variety of system properties \cite{A.Ol2014Minimal,T2016On,clark2014minimizing,zhang2017sensor}. Depending on the different design purposes, some I/O selection problems are shown to be NP-hard \cite{A.Ol2014Minimal} or polynomially solvable \cite{S.Pe2016A}, and some share certain nice properties, such as being submodular which enables easy approximations \cite{clark2014minimizing}.


The past decade has also witnessed a growing research interest in I/O selections for {\emph{structured systems}} \cite{Y.Y.2011Controllability,ramos2020structural}, i.e., systems whose state-space representations are described by structured matrices \cite{generic}. Structured systems can often be represented by graphs, and have the potential to describe the interconnection structure of large-scale networked systems \cite{Y.Y.2011Controllability}. In particular, some properties defined on structured systems are generic in the sense that almost all realizations of a structured system share the same said properties. Controllability is such a property, and the corresponding notion is {\emph{structural controllability}} \cite{Lin_1974}.

 Among the related problems, the problems of optimally adding/selecting (resp. deleting) actuators or interconnection links to make the resulting system {\emph{structurally controllable}} (or structurally uncontrollable) have been extensively explored \cite{S.Pe2016A,A_Olshevsky_2015,pequito2016minimum,Y_Zhang_2017,zhang2019minimal,dey2018minimum}.  Particularly, \cite{S.Pe2016A} gives the first polynomial time algorithm for selecting the minimum number of state variables to be affected by dedicated inputs to ensure structural controllability.\footnote{A dedicated input is the input that actuates only one state variable.} \cite{A_Olshevsky_2015} extends the previous result by restricting that some state variables are forbidden to be actuated and providing a faster algorithm. Latter, \cite{pequito2016minimum} shows the polynomial solvability of the problem of selecting state variables with the minimum total cost to be affected by inputs to ensure structural controllability, where actuating a state variable incurs a non-negative cost that is {\emph{independent}} of the inputs. So far, all the polynomially solvable cases reported in \cite{S.Pe2016A,A_Olshevsky_2015,pequito2016minimum} belong to the dedicated input case. Only very little was known about the complexity status for the non-dedicated constrained input case, except for some cases reported in \cite{zhang2019minimal,Y_Zhang_2017,dey2018minimum} where the corresponding problems could be {\emph{trivially}} reduced to the minimum (cost) spanning arborescence problems or the minimum cost maximum matching problems. \footnote{\label{footnote1}We remark that when there is no cost imposed on the input links and no constraint on the connectivity between each input and all state variables, the optimal (sparsest) input selection problems are trivially equivalent to the corresponding dedicated input cases \cite{S.Pe2016A}.}

 In this paper, we make an attempt towards the non-dedicated constrained input case and provide an alternatively algebraic method for it. More precisely, given an autonomous system and a {\emph{constrained}} input configuration, where whether an input can directly actuate a state variable, as well as the corresponding (possibly different) cost, is prescribed, we consider three related cost-sparsity induced optimal input selection problems: selecting the minimum number of input links, selecting the minimum (total) cost of input links, and selecting the input links with the smallest cost while their cardinality does not exceed a prescribed number, all for ensuring structural controllability. To the best of our knowledge, the third problem has not been considered before, and no polynomial time algorithms have been reported for all the three problems in the non-dedicated input case, except for some trivial cases (c.f. \cite{zhang2019minimal,Y_Zhang_2017,dey2018minimum}).

 In this paper, as our first contribution, we show these problems can be formulated as equivalent integer linear programming (ILP) problems by suitably choosing the decision variables. Although ILPs are usually NP-hard, we reveal that the corresponding constraint matrices of these ILPs are {\emph{totally unimodular}}, under a weak constraint on the input configuration which contains the dedicated input one as a special case, the so-called source strongly-connected component grouped input constraint (and this defines the most possible class of systems with such a property). {\emph{Total unimodularity (TU)}} is an important property for combinatorial problems \cite{lawler2001combinatorial,hoffman201613}, but has not yet been revealed for the input selection problems as far as we know. This inherent structure allows us to solve those ILPs efficiently by simply removing the integer constraints and solving the linear programming (LP) relaxations. Hence, as our second contribution, we provide a unifying LP-based method with polynomial time complexity towards the three problems, which also gives an algebraic, rather than graph-theoretic proof for the polynomial solvability of these problems for a wide variety of non-dedicated input constraints.
 Furthermore, thanks to the TU structure, it is revealed that the considered problems are strongly polynomially solvable under the addressed condition, meaning there are algorithms that can solve them in polynomial time that is independent of the size of the costs and cardinality bounds.

 The rest of this paper is organized as follows. Section \ref{problem-formulation} gives the problem formulations, and Section \ref{preliminary} provides some preliminaries in graph theory and structured systems. Section \ref{main-ILP} presents ILP formulations of the addressed problems, while Section \ref{main-TU} deals with their TU properties and efficient solvability. Section \ref{example} provides an illustrative example. The last section concludes this paper.



{\bf Notations and terminologies:} For two vectors $a$ and $b$, $a\le b$ means $a_i\le b_i$ entry-wisely. For an optimization problem $\min \{\varphi(x): x\in \Lambda\}$, $\Lambda$ is the feasible region, $x\in \Lambda$ is a feasible solution, the minimum of the objective $\varphi(x)$ on $x\in \Lambda$ is called the optimal (objective) value, or optimum, while the $x$ for which the optimum is attained is called an optimal solution. An optimal solution $x$ is called the integral optimal solution, if $x$ is integral. $1_{n\times m}$ denotes the $n\times m$ matrix with all entries $1$.

\section{Problem formulations} \label{problem-formulation}
Consider a linear-time invariant system
\begin{equation}
\dot x(t)=\tilde A x(t) + \tilde B u(t),
\end{equation}
where $x(t)\in {\mathbb R}^n$, $u(t)\in {\mathbb R}^m$ are the state variables and inputs, and $\tilde A\in {\mathbb R}^{n\times n}$, $\tilde B\in {\mathbb R}^{n\times m}$ are the state transition matrix and input matrix, respectively.

Let $A$ and $B$ be structured matrices that characterize the sparsity patterns of $\tilde A$ and $\tilde B$, that is, $A_{ij}=0$ (resp. $B_{ij}=0$) implies $\tilde A_{ij}=0$ (resp. $\tilde B_{ij}=0$), for all $1\le i,j \le n$ (resp. $1\le i \le n, 1\le j \le m$). We may use $\{0,*\}^{n_1\times n_2}$ to denote the set of all structured matrices with the dimension $n_1\times n_2$, in which $0$ denotes the fixed zero entries, and $*$ the entries that can take values freely. For a structured matrix $M\in \{0,*\}^{n_1\times n_2}$, ${\cal S}(M)$ denotes the set of its realizations, i.e., ${\cal S}(M)=\{\tilde M\in {\mathbb R}^{n_1\times n_2}: \tilde M_{ij}=0 \ {\rm if} \ M_{ij}=0\}$. $(A,B)$ is said to be structurally controllable, if there exists $\tilde A\in {\cal S}(A)$ and  $\tilde B \in {\cal S}(B)$, so that $(\tilde A, \tilde B)$ is controllable. It is well-known that controllability is a generic property, in the sense that if $(A,B)$ is structurally controllable, then almost all of its realizations are controllable.


Given $B \in \{0,*\}^{n\times m}$, let ${\cal N}(B)$ be the set of $*$ entries in $B$, i.e., ${\cal N}(B)=\{(i,j): B_{ij}=*\}$. Define the set of $n\times m$ structured matrix as
$${\cal K}(B)=\{B': {\cal N}(B')\subseteq {\cal N}(B)\}.$$
We say $B$ is {\emph{dedicated}}, if each column of $B$ has at most one nonzero entry. Assign a non-negative cost $w_{ij}$ to each nonzero entry $B_{ij}$ of $B$, representing the cost of actuating the $i$th state variable using the $j$th input. Let $||B||_w$ be the sum of costs of all nonzero entries (corresponding to input links) in $B$, i.e., $$||B||_{w}=\sum \nolimits_{(i,j)\in {\cal N}(B)} w_{ij}.$$Let $||B||_0$ be the number of nonzero entries (i.e., sparsity) of $B$. With notations above, we consider the following three optimal input selection problems:

Problem ${\cal P}_1$:  constrained sparsest input selection
\begin{equation}\begin{array}{l}
\min \limits_{B'\in {\cal K}(B)} ||B'||_0 \tag{${\cal P}_1$}\label{A1} \\
{\rm s.t.} \ (A,B') \ {\rm structurally \ controllable}
\end{array}\end{equation}

Problem ${\cal P}_2$: constrained minimum cost input selection
\begin{equation}\begin{array}{l}
\min \limits_{B'\in {\cal K}(B)} ||B'||_w \tag{${\cal P}_2$}\label{A2}  \\
{\rm s.t.} \ (A,B')\ {\rm structurally \ controllable}
\end{array}\end{equation}

Problem ${\cal P}_3$: minimum cost $k$-sparsity input selection ($k$ is given)
\begin{equation}\begin{array}{l}
\min \limits_{B'\in {\cal K}(B)} ||B'||_w \tag{${\cal P}_3$}\label{A3} \\
{\rm s.t.} \ (A,B')\ {\rm structurally \ controllable} \\
    \ \ \ \ \ ||B'||_0\le k
\end{array}\end{equation}

The problems above all assume independence of selecting each available input link. As their names suggested, problem ${\cal P}_1$ seeks to select the sparsest input matrix from $B$, problem ${\cal P}_2$ aims to select the input matrix from $B$ with the smallest total cost of its input links, while problem ${\cal P}_3$ intends to find the input matrix from $B$ with a bound on its sparsity and with the total cost of input links as small as possible, all to ensure structural controllability. It is remarkable that problem ${\cal P}_3$ is more general than the problem $(2)$ discussed in \cite{pequito2016minimum}, the latter of which is, in fact, a special case of the former by setting $k$ to be the optimal value of problem ${\cal P}_1$ with $B$ being an $n\times n$ matrix full of nonzero entries (see footnote 2 for the unconstrained case), and indeed can also be formulated as problem ${\cal P}_2$ by {\emph{reassigning the link costs}} (see Remark \ref{remark-sparest}). Problem ${\cal P}_3$ may be desirable, for example, in a communication network where the activation of new communication links may be expensive, while different links may have different operating budgets/communication qualities (characterized by the cost $[w_{ij}]$), the goal is to select the input configuration with the size of active input communication links not exceeding a prescribed number to ensure system controllability, while making the total budgets/qualities as small/good as possible. The example in Section \ref{example} will highlight the differences among these problems. 



Throughout, the following assumptions are adopted:

 \begin{assumption} \label{sc-assump}
$(A,B)$ is structurally controllable.
\end{assumption}
\begin{assumption} \label{cost-assump}
Let $w_{\min}=\min \nolimits_{(i,j)\in {\cal N}(B)} w_{ij}$ and $w_{\max}=\max \nolimits_{(i,j)\in {\cal N}(B)} w_{ij}$. Assume that $0\le w_{\min}\le w_{\max}< \infty$.
\end{assumption}


Assumption \ref{sc-assump} is necessary for the feasibility of problems ${\cal P}_1, {\cal P}_2$, and ${\cal P}_3$. To ensure problem ${\cal P}_3$ is feasible, one additional requirement is that $k$ should be no less than the optimal value of problem ${\cal P}_1$ (denoted by ${N}^\star_{{\cal P}_1}$). Assumption \ref{cost-assump} is general enough to meet all practical designs. Say, if all input links are of equal cost, then $w_{\max}=w_{\min}>0$; if some input links already exist, then $w_{\min}=0$; in the most general case, different input links may have heterogeneous costs. When each input link has a uniform cost, problem ${\cal P}_2$ is equivalent to problem ${\cal P}_1$. If $k\ge n$, problem ${\cal P}_3$ is equivalent to problem ${\cal P}_2$ (c.f. Theorem \ref{sc-theory}). As mentioned earlier, in the non-dedicated input case, no combinatorial algorithms have been reported for all the three problems (except for some trivial cases in \cite{zhang2019minimal,Y_Zhang_2017,dey2018minimum}).

\section{Preliminaries} \label{preliminary}
This section introduces some preliminaries in graph theory and structured systems. 

A directed graph (digraph for short) is denoted by $G=(V,E)$, with $V$ the vertex set and $E$ the edge set. A path in a digraph is a set of ordered edges, in which the terminal vertex of the preceding edge is the starting vertex of the successive edge.  A digraph is strongly connected, if for any pair of its vertices, there is a path from each of them to the other. A strongly-connected component (SCC) of a digraph is its subgraph that is strongly connected, and no edges or vertices can be included in this subgraph without breaking its property of being strongly connected. A bipartite graph, which often reads $G=(V_L,V_R,E_{RL})$, is a graph whose vertices can be partitioned into two parts $V_L$ and $V_R$, such that all its edges $E_{RL}$ have end vertices in both parts.  A matching of a bipartite graph is a set of edges among which any two do not share a common end vertex. A vertex is matched with respect to a matching, if it is contained in this matching. The maximum matching is the matching with as many edges as possible. 

Given $A\in \{0,*\}^{n\times n}$, $B\in \{0,*\}^{n\times m}$, the state digraph  is ${\cal G}(A)=(X,E_{A})$, in which $X=\{x_1,...,x_n\}$ is the set of state vertices, $E_{A}=\{(x_j,x_i): A_{ij}\ne 0\}$ is the set of state edges. The system digraph is ${\cal G}(A,B)=(X\cup U, E_{A}\cup E_{B})$, where the input vertices $U=\{u_1,...,u_m\}$, the input links (edges) $E_{B}=\{(u_i,x_j): B_{ji}\ne 0\}$. Corresponding to ${\cal G}(A,B)$, the bipartite graph associated with $(A,B)$ is defined as ${\cal B}(A,B)=(X_L,U\cup X_R, E_{XX}\cup E_{UX})$, in which $X_L=\{x^L_1,...,x^L_n\}$, $X_R=\{x_1^R,...,x_n^R\}$, $U=\{u_1,...,u_m\}$, $E_{XX}=\{(x^R_j,x^L_i): A_{ij}\ne 0\}$, and $E_{UX}=\{(u_j,x^L_i): B_{ij}\ne 0\}$.

Suppose ${\cal G}(A)$ can be decomposed into $n_c$ SCCs,  $1\le n_c \le n$, and the $i$th SCC has a vertex set $X_i\subseteq X$ ($1\le i \le n_c$). An SCC is called a {\emph{source SCC}}, if there is no incoming edge to vertices in this SCC from other SCCs in ${\cal G}(A)$; otherwise, we call it a {\emph{non-source SCC}}.   Suppose there are $r$ source SCCs in ${\cal G}(A)$, with their indices being ${\cal I}=\{1,...,r\}$, $1\le r \le n_c$. For each $i\in {\cal I}$, let $X_i^L=\{x^L_j\in X_L: x_j\in X_i\}$, and define $E_i=\{(u,x)\in E_{UX}: x\in X_i^L, u\in U\}$ as the set of input links between $U$ and $X^L_i$ in ${\cal B}(A,B)$. A state vertex $x_i\in X$ is said to be {\emph{input-reachable}}, if there is a path starting from an input vertex $u\in U$ to $x_i$ in ${\cal G}(A,B)$. With a little abuse of terminology, if each vertex of $X_i$ is input-reachable in ${\cal G}(A,B)$, we just say $X_i^L$ is input-reachable in ${\cal B}(A,B)$. 

 \begin{theorem}[\cite{generic}] \label{sc-theory} $(A,B)$ is structurally controllable, if and only if: i) every state vertex $x_i\in X$ is input-reachable, and ii) there is a maximum matching in ${\cal B}(A,B)$ such that every $x_i^L\in X_L$ is matched.
 \end{theorem}

By the definition of input-reachability, it is obvious condition i) of Theorem \ref{sc-theory} is equivalent to that, $E_i\ne \emptyset$ for each $i\in {\cal I}$.
\section{ILP formulations of ${\cal P}_1$, ${\cal P}_2$, and ${\cal P}_3$} \label{main-ILP}
In this section, we formulate problems ${\cal P}_1$, ${\cal P}_2$, and ${\cal P}_3$ as some equivalent ILPs.  Before introducing our ILP formulations, we discuss the {\emph{essential difficulty}} in extending the graph-theoretic algorithms for problems ${\cal P}_1, {\cal P}_2$ in \cite{S.Pe2016A, pequito2016minimum} from the dedicated input case to the non-dedicated one. From Theorem \ref{sc-theory}, any subset of $E_{UX}$ making $(A,B')$ structurally controllable can be divided into two parts $E_{\rm mat}$ and $E_{\rm rea}$, so that the addition of $E_{\rm mat}$ to $(X_L,U\cup X_R, E_{XX})$ makes the resulting $(X_L, U\cup X_R, E_{XX}\cup E_{\rm mat})$ have a maximum matching that matches $X_L$, and the addition of $E_{\rm rea}$ makes the obtained $(X_L,U\cup X_R, E_{XX}\cup E_{\rm rea})$ have a nonempty $E_i$ for each $i\in {\cal I}$. Although both parts with the minimum cardinality/cost can be polynomially determined via the respective graph-theoretic algorithms: i.e., via the minimum cost maximum matching algorithm for $E_{\rm mat}$ and via the minimum (cost) spanning arborescence algorithm for $E_{\rm rea}$ (see \cite{zhang2019minimal,Y_Zhang_2017,dey2018minimum} for details), respectively, the challenge is that $E_{\rm mat}$ and $E_{\rm rea}$ might overlap, thus making their union not necessarily optimal. The essential idea of the graph-theoretic methods in \cite{S.Pe2016A, pequito2016minimum} is to find the maximal `intersection' between these two sets. To this end, by introducing some slack variables to $(X_L,X_R,E_{XX})$, they construct a weighted bipartite graph, and for a weighted maximum matching $E_s$ of it, an optimal input solution is obtained by selecting {\emph{dedicated inputs}} to those state vertices that are not matched by $E_s\cap E_{XX}$, and adding additional inputs to make every state vertex input-reachable. In the dedicated input case, selecting an input link will not affect the ability of the subsequent input links w.r.t. the matching function (in a feasible solution $E_{\rm mat}\cup E_{\rm rea}$ to problem ${\cal P}_1$ or ${\cal P}_2$, we say an input link $e$ serves the matching function if $e\in E_{\rm mat}$). However, in the non-dedicated input case, things are different since multiple input-links may be incident to the same input vertex. This makes extending the graph-theoretic algorithms in \cite{S.Pe2016A, pequito2016minimum} to the non-dedicated input case nontrivial. 

\vspace{-0.4em}
 In our ILP formulations for the general input case, however, we do not intend to figure out such intersection; instead, we directly adopt the corresponding cost functions in problems ${\cal P}_1$, ${\cal P}_2$, and ${\cal P}_3$ as our objectives. The key to the ILP formulations is the introduction of binary variables $y=\{y_{uv}: (u,v)\in E_{XX}\cup E_{UX}\}$ and $z=\{z_i: i\in {\cal I}\}$, where $y_{uv}=1$ indicates edge $(u,v)$ is in a specific maximum matching $E_s$ of ${\cal B}(A,B)$, $y_{uv}=0$ means the contrary; and $z_i=1$ means $X^L_i$ is input-reachable after adding the edges $E_s\cap E_{UX}$ to $(X_L, X_R\cup U, E_{XX})$, $z_i=0$ means the contrary. The  variables $y$ and $z$ will be the decision variables to our ILP formulations, presented formally in the following theorem.

\vspace{-0.4em}
\begin{theorem}\label{ILP-formulation} Under Assumptions \ref{sc-assump}-\ref{cost-assump}, problem ${\cal P}_i$ is equivalent to the following ILP ${\cal P}^{\rm ILP}_i$ (i.e., their optimal objective values are equal), for $i=1,2,3$,  respectively:
\begin{align}
\min_{y,z} \quad & \sum \nolimits_{(u,v)\in E_{UX}} y_{uv} + |{\cal I}|- \sum \nolimits_{i\in {\cal I}} z_i \tag{${\cal P}^{\rm ILP}_1$}\label{ILP1} \\
{\rm{s.t.}}\quad & \sum \nolimits_{(u,v)\in E_{XX}\cup E_{UX}} y_{uv} = 1, \forall v\in X_L \label{C1} \\
& \sum \nolimits_{(u,v)\in E_{XX}\cup E_{UX}} y_{uv} \le 1, \forall u \in X_R\cup U \label{C2}\\
& z_i \le \sum \nolimits_{(u,v)\in E_i} y_{uv}, \forall i\in {\cal I} \label{C3} \\
& y_{uv}\in \{0,1\}, \forall (u,v)\in E_{XX}\cup E_{UX} \label{C4} \\
& z_{i}\in \{0,1\}, \forall i\in {\cal I}. \label{C5}
\end{align}
\begin{align}
\min_{y,z} \quad & \sum \nolimits_{(u,v)\in E_{UX}} w_{uv}y_{uv} + \sum \nolimits_{i\in {\cal I}} (1-z_i)w_i^{\min} \tag{${\cal P}^{\rm ILP}_2$}\label{ILP2} \\
{\rm{s.t.}}\quad & (\ref{C1}), (\ref{C2}), (\ref{C3}), (\ref{C4}), {\rm and}\ (\ref{C5}) \label{C6}
\end{align}
\begin{align}
\min_{y,z} \quad & \sum \nolimits_{(u,v)\in E_{UX}} w_{uv}y_{uv} + \sum \nolimits_{i\in {\cal I}} (1-z_i)w_i^{\min} \tag{${\cal P}^{\rm ILP}_3$}\label{ILP3} \\
{\rm{s.t.}}\quad & \sum \nolimits_{(u,v)\in E_{UX}} y_{uv} + |{\cal I}|- \sum \nolimits_{i\in {\cal I}} z_i\le k \ \label{C7} \\
\quad & (\ref{C1}), (\ref{C2}), (\ref{C3}), (\ref{C4}), {\rm and}\ (\ref{C5}) \ \label{C8}
\end{align}where $w_{u_j,x_i^L}=w_{ij}$ for $(u_j,x_i^L)\in E_{UX}$, and $w_i^{\min}= \min \nolimits_{(u,v)\in E_i} w_{uv}$, i.e., the minimum cost of input links incident to $X^L_i$, for each $i\in {\cal I}$.
Besides, with Assumptions \ref{sc-assump}-\ref{cost-assump}, for any $k\in {\mathbb N}$, problem ${\cal P}_3$ is feasible, if and only if the ILP ${\cal P}^{\rm ILP}_3$ is.
\end{theorem}

\begin{proof} We first focus on ${\cal P}_1^{\rm ILP}$. Let $E_s=\{(u,v)\in E_{XX}\cup E_{UX}: y_{uv}=1, (y,z) \ {\rm subject\ to} \ (\ref{C1})-(\ref{C5})\}$. Constraint (\ref{C1}) means every vertex of $X_L$ should be an end vertex of exactly one edge in $E_s$, and constraint (\ref{C2}) means each vertex of $X_R\cup U$ can be the end vertex of at most one edge in $E_s$. Therefore, constrains (\ref{C1}) and (\ref{C2}) make sure $E_s$ is a matching of ${\cal B}(A,  B)$ that matches $X_L$. Moreover, to minimize the objective function, $z_i$ subject to (\ref{C3}) and (\ref{C5}) should take the value $z_i=\min \{\sum \nolimits_{(u,v)\in E_i} y_{uv}, 1\}$, for each $i\in {\cal I}$; otherwise, by changing the corresponding $z_i$ from $0$ to $1$, the constraints are fulfilled while the objective value can decrease. That is, if $X_i^L$ is input-reachable in $(X_L,X_R\cup U, E_s\cup E_{XX})$, then $z_i=1$; otherwise, $z_i=0$. Hence, to make $X_i^L$ input-reachable for all $i\in {\cal I}$, the minimum number of input links that need to be added to $(X_L,X_R\cup U, E_s\cup E_{XX})$ is $|{\cal I}|-\sum \nolimits_{i\in {\cal I}}z_i$, and adding one element arbitrarily from $E_i$ for each $i$ with $z_i=0$ is feasible. Therefore, optimizing the objective of ${\cal P}_1^{\rm ILP}$ will obtain the optimal solution to ${\cal P}_1$.  On the other hand, according to Theorem \ref{sc-theory} and following a similarly reversed analysis, with Assumption \ref{sc-theory}, it turns out that any optimal solution to problem ${\cal P}_1$ should correspond to a $(y^{\star},z^{\star})$ that optimizes ${\cal P}_1^{\rm ILP}$.


We now consider ${\cal P}_2^{\rm ILP}$. With the notations defined above, after adding $E_s$ to $(X_L,X_R\cup U, E_{XX})$, to make every $X_i^L$ with $z_i=0$ input-reachable while incurring the minimal cost, it suffices to select the input link incident to $X_i^L$ with the minimal cost, i.e., the input link in $E_i$ with the cost $w_i^{\min}$. On the other hand, to minimize the objective function, for an $i\in {\cal I}$ with $w_i^{\min}>0$, $z_i$ subject to (\ref{C7}) and (\ref{C8}) must take $z_i=\min\{\sum \nolimits_{(u,v)\in E_i} y_{uv},1\}$, as otherwise one can always change the respective $z_i$ from $0$ to $1$, so that constraints (\ref{C7}) and (\ref{C8}) are fulfilled, while the objective value decreases (note if $w_i^{\min}=0$, the objective value remains still whenever $z_i$ takes $0$ or $\min\{\sum \nolimits_{(u,v)\in E_i} y_{uv},1\}$).   Therefore, the optimal value of ${\cal P}_2^{\rm ILP}$ is equal to that of problem ${\cal P}_2$.

Consider ${\cal P}_3^{\rm ILP}$.  Based on the analysis for ${\cal P}_1^{\rm ILP}$, for a fixed matching $E_s$ defined therein, the left-hand side of (\ref{C7}) is the minimal possible number of input links associated with $E_s$ that can make the original system structurally controllable. In other words, constraint (\ref{C7}) (along with constraint (\ref{C8}), which ensures the matching that matches $X_L$ exists) ensures the sparsity of the feasible input matrix for problem ${\cal P}_3$ does not exceed $k$. Hence, ${\cal P}_3$ is feasible, if and only if ${\cal P}_3^{\rm ILP}$ is.   Then, following the analysis for ${\cal P}_2^{\rm ILP}$, it turns out the optimal value of ${\cal P}_3^{\rm ILP}$ is equal to that of problem ${\cal P}_3$. 
\end{proof}

The following theorem states how to recover an optimal solution to problem ${\cal P}_i$ from the corresponding ${\cal P}_i^{\rm ILP}$ ($i=1,2,3$).

\begin{theorem}\label{solution} Suppose Assumptions \ref{sc-assump}-\ref{cost-assump} hold. Let $(u_i^{\min},x_i^{\min})=\arg \min \nolimits_{(u,x)\in E_i} w_{ux}$, for each $i\in {\cal I}$.\footnote{If each input edge in $E_i$ is of equal cost, $(u_i^{\min},x_i^{\min})$ can be any element of $E_i$. In other words, for problem ${\cal P}_1$, $(u_i^{\min},x_i^{\min})$ can be arbitrarily selected. } Let $(y^{\star},z^{\star})$ be an optimal solution to ${\cal P}_i^{\rm ILP}$ ($i=1,2,3$). Define
 $$ \begin{array}{l} E_{\rm mat}^{\star}=\{(u,v)\in E_{UX}: y^{\star}_{uv}=1\}, \\
 E_{\rm rea}^{\star}=\{(u_i^{\min},x_i^{\min}): z^{\star}_i=0, i\in {\cal I}\}.
 \end{array}
 $$
Then, $E_{\rm mat}^{\star}\cup E_{\rm rea}^{\star}$ is the set of input links of an optimal solution to problem ${\cal P}_i$ ($i=1,2,3$).
\end{theorem}

\begin{proof} The statement follows directly from Theorem \ref{sc-theory} and the proof of Theorem \ref{ILP-formulation}.
\end{proof}

\begin{remark} Note in ${\cal P}^{\rm ILP}_1$, ${\cal P}^{\rm ILP}_2$, and ${\cal P}^{\rm ILP}_3$, the relation $z_i=\min\{\sum\nolimits_{(u,v)\in E_i}y_{uv},1\}$ (for $w_i^{\min}>0$) results from optimizing the objective functions, rather than the feasible regions. Without Assumption \ref{sc-assump}, it may happen that problem ${\cal P}_i$ ($i=1,2$, or $3$) is not feasible, but ${\cal P}_i^{\rm ILP}$ is. The advantage of ${\cal P}^{\rm ILP}_3$ lies in that, with Assumption \ref{sc-assump}, we need not pre-check whether $k$ is feasible for problem ${\cal P}_3$; instead, ${\cal P}^{\rm ILP}_3$ will tell.
\end{remark}

\begin{remark} The optimal solution to problem ${\cal P}_1$ may not be unique. In particular, there is much flexibility in choosing $E^{\star}_{\rm rea}$ for problem ${\cal P}_1$.  Theorem \ref{solution} provides a basic solution to problem ${\cal P}_1$, and from it, other possible solutions may be found by some graph-theoretic transformations.
\end{remark}

\begin{remark} \label{remark-sparest}
Finding the sparsest input matrix from ${\cal K}(B)$ while incurring the total cost as small as possible for ensuring structural controllability (c.f. problem (2) in \cite{pequito2016minimum,dey2018minimum}, referred to as ${\cal P}_4$) can be alternatively formulated as the following ILP, given ${w_{\max}}>0$:
\begin{align}
\min_{y,z} \quad & \sum \nolimits_{(u,v)\in E_{UX}} w_{uv}y_{uv} + \sum \nolimits_{i\in {\cal I}} (1-z_i)w_i^{\min} \tag{${\cal P}^{\rm ILP}_4$}\label{ILP4} \\
\quad \quad & + \gamma(\sum \nolimits_{(u,v)\in E_{UX}} y_{uv} + |{\cal I}|- \sum \nolimits_{i\in {\cal I}} z_i)  \\
{\rm{s.t.}} \quad & (\ref{C1}), (\ref{C2}), (\ref{C3}), (\ref{C4}), {\rm and}\ (\ref{C5}) \ \label{C8add}
\end{align}
where $\gamma=nw_{\max}$ is to penalize the sparsity of the solution, such that for any feasible solution with the sparsity larger than $N^{\star}_{{\cal P}_1}$, its possible decrease in the total link cost will not exceed the increase caused by the sparsity penalty. Note ${\cal P}_{4}^{\rm ILP}$ can be formulated as ${\cal P}_2^{\rm ILP}$ by redefining the cost $w_{uv}\doteq w_{uv}+\gamma$, $\forall (u,v)\in E_{UX}$, meaning ${\cal P}_4$ is indeed a special case~of~${\cal P}_2$.
\end{remark}

\section{LP relaxations and total unimodularity of ${\cal P}_1$, ${\cal P}_2$, and ${\cal P}_3$} \label{main-TU}
Although we have formulated problems ${\cal P}_i|_{i=1}^3$ as ILPs, it is less favorable unless those ILPs can be solved efficiently. It is known that ILP is usually NP-hard (for example, the set cover problem) \cite{lawler2001combinatorial}. Nevertheless, by proving TU of the constraint matrices of the ILPs,  we will show ${\cal P}^{\rm ILP}_i|_{i=1}^3$ could be solved in polynomial time simply via their LP relaxations, and even in strongly polynomial time, under a wide condition on $B$ that contains almost all the existing known nontrivial conditions with which problems ${\cal P}_1$ and ${\cal P}_2$ are reportedly polynomially solvable.  
\vspace{-1em}
\subsection{Total unimodularity}
\begin{definition}[Total unimodularity \cite{lawler2001combinatorial}, TU] A matrix $M$ is TU if every square submatrix has determinant $0, +1$, or $-1$.
\end{definition}

\begin{lemma}[\cite{hoffman201613}] \label{extreme-point}
Let $M$ be an $p\times q$ TU matrix. Then every extreme point (vertex) of the following polyhedron is integral for any vectors $b$ of integers\footnote{For notions of polyhedron and extreme point, we refer readers to \cite{lawler2001combinatorial,schrijver1998theory}.}:
\begin{equation}\label{standard-LP}\{x\in {\mathbb R}^{q}: Mx\le b, x\ge 0\}.\end{equation}
\end{lemma}


 According to the fundamental theorem of linear programming, every optimal solution of an LP problem (if exists) is either an extreme point of its feasible polyhedron (region), or lies on a face of optimal solutions (i.e., every convex combination of its optimal solutions) \cite{schrijver1998theory}. Based on this and Lemma \ref{extreme-point}, if the ILP in its standard form $\min \{c^\intercal x: Mx\le b, x\ge 0, x\in {\mathbb N}^q\}$ has a $p\times q$ TU constraint matrix $M$, then its LP relaxation by removing the integral constraint on $x$, i.e., $\min \{c^\intercal x: Mx\le b, x\ge 0\}$ yields an integral optimal solution \citep[Chap. 4.12]{lawler2001combinatorial} whenever the optimum exists and is finite, for any vector $b$ of integers and all rational $c$. That's to say, the original ILP can be solved efficiently by standard LP algorithms. It also follows that the original ILP is polynomial-time solvable, since the respective LP is \citep[Theorem 16.2]{schrijver1998theory}.\footnote{It is remarkable that some LP solvers may return a non-integral optimal solution (this happens when there exist an integral and a non-integral solution which are both optimal). In this case, an integral optimal solution can always be found from the non-integral one via some standard manipulations in polynomial time. For example, \citep[Corollary 5.3b]{schrijver1998theory} accomplishes this using the Hermite normal form, which incurs complexity approximately $\tilde O(d^w)$, with $w<2.373$ being the exponent of matrix multiplication, and $d$ the number of decision variables; see \citep[Corollary 5.3b, Theorem 16.2]{schrijver1998theory} for details.} 

We introduce the following constraint on $B$, termed the {\emph{source-SCC grouped input constraint}}. As shown in the subsequent Propositions \ref{unimodular}-\ref{unimodular-addcardinality} and Example \ref{asp-not-satisfied}, this constraint defines the most possible class of systems whose corresponding constraint matrices of ${\cal P}^{\rm ILP}_i|_{i=1}^3$ are TU.


\begin{assumption} \label{broader-assumption} (Source-SCC grouped input constraint) Given $(A,B)$, assume in ${\cal G}(A,B)$ no input vertex can simultaneously actuate two state vertices that come from different source SCCs, or one is from a source SCC and the other a non-source SCC.
\end{assumption} 

\begin{figure} \centering
\subfigure[] { \label{fig:a}
\includegraphics[width=0.21\columnwidth]{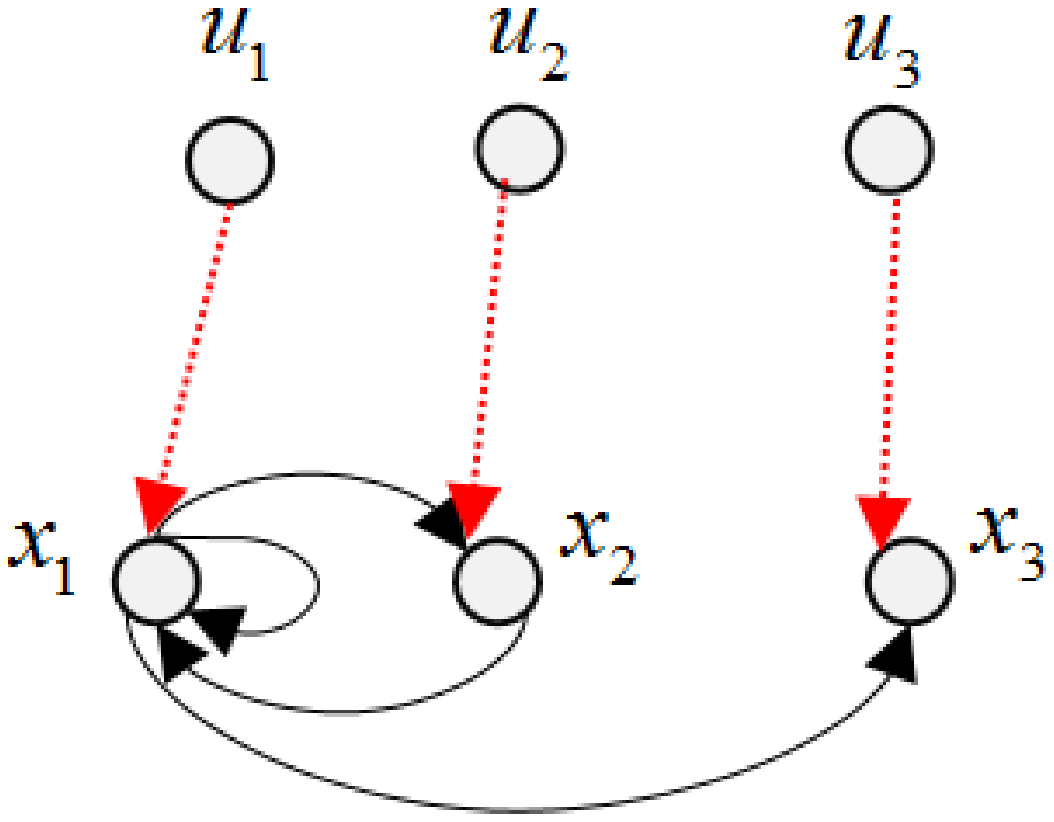}
}
\subfigure[] { \label{fig:b}
\includegraphics[width=0.21\columnwidth]{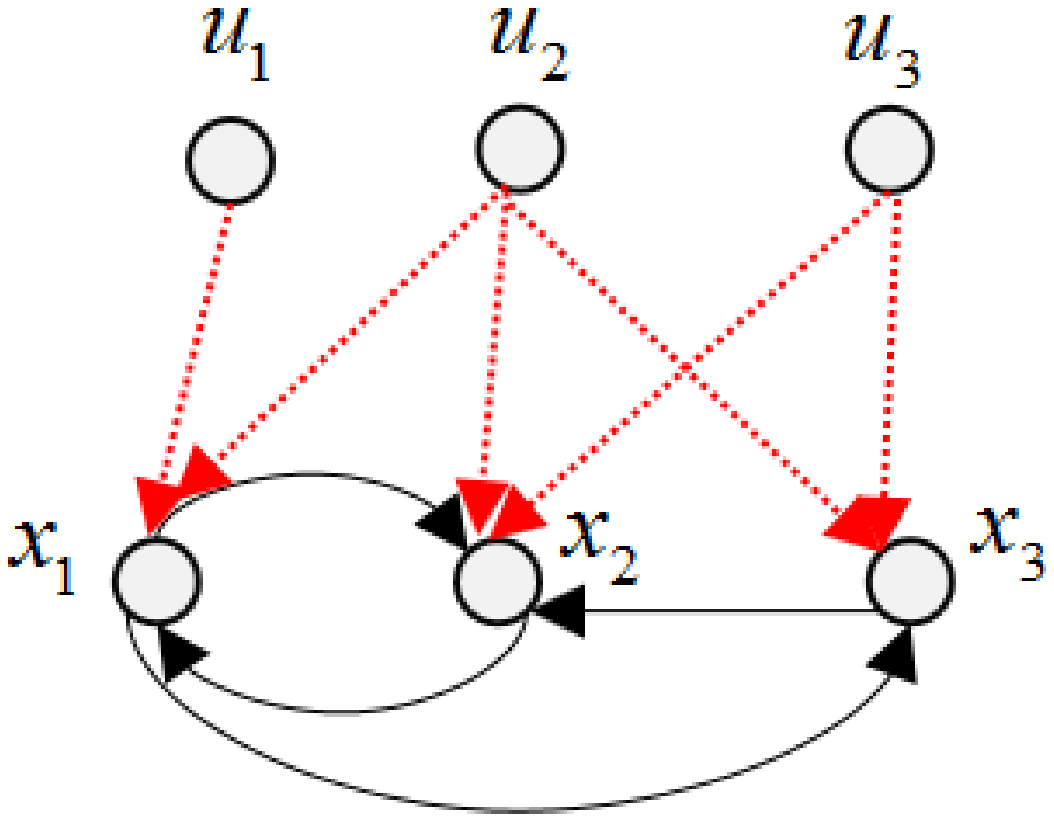}
}
\subfigure[] { \label{fig:c}
\includegraphics[width=0.273\columnwidth]{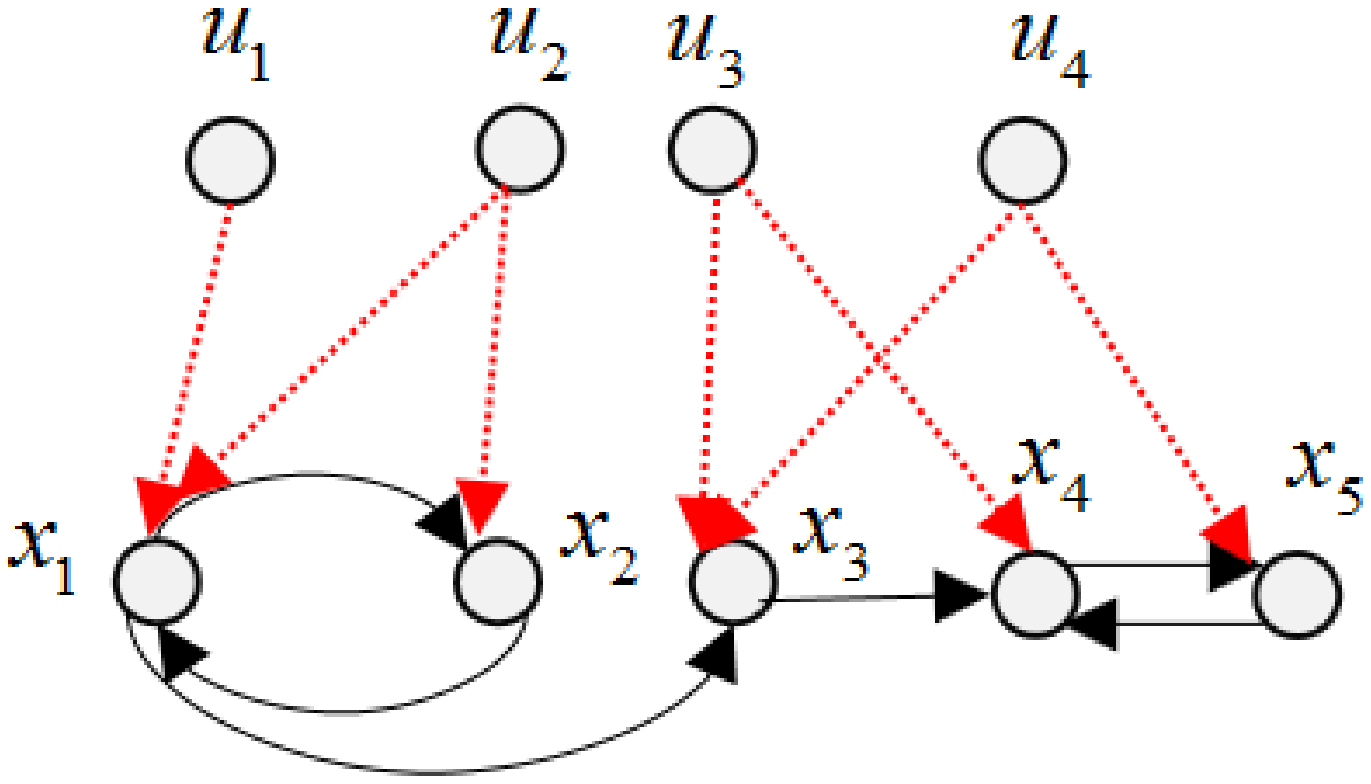}
}
\caption{Examples that satisfy the source-SCC grouped input constraint. (a): dedicated input case; (b): strongly connected case;  (c): general case. }
\label{assump3-example}
\end{figure}

Note Assumption \ref{broader-assumption} does {\emph{not}} impose constraints on how each input vertex connects with vertices within the same source SCC, or how each input vertex connects with vertices not belonging to the source SCCs (thus, one input vertex can connect with multiple non-source SCCs). Particularly, the dedicated input case falls into Assumption \ref{broader-assumption}. The case where ${\cal G}(A)$ is strongly connected automatically satisfies Assumption \ref{broader-assumption}. See Fig. \ref{assump3-example} for a wide variety of scenes where this input constraint is met.  It seems Assumption \ref{broader-assumption} characterizes a two-layer input structure, in which the first layer consisting of all source SCCs, and the second one all non-source SCCs. One input signal cannot directly actuate two state variables that are in different layers. In the second layer, every input signal can arbitrarily actuate the corresponding state variables therein, while in the first layer, each source SCC somehow has a high priority of autonomy such that it has its own inputs. Such a hierarchical input configuration may be often in social networks, political networks, influence networks, etc., where layered/hierarchical structures often emerge \cite{liu2012control}. We note even with this constraint, there might be some obstacles in extending the graph-theoretic algorithms in \cite{S.Pe2016A,A_Olshevsky_2015,pequito2016minimum} towards these problems.

To show the TU property of ${\cal P}_i^{\rm ILP}|_{i=1}^3$, in what follows, we characterize the TU of two augmented incidence matrices of a general bipartite graph satisfying a condition resembling Assumption \ref{broader-assumption},  which might be of independent interest. The proofs are postponed to Section \ref{proof-TUM}.


{\linespread{1.1} \selectfont
\begin{proposition}\label{unimodular} Suppose in a bipartite graph $G=(X_L,X_R\cup U,E)$ (not necessarily corresponding to a structured system), $X_L$ and $U$ are partitioned into $r+1$ ($r\ge 0$) disjoint subsets $X^{C_1}_L,\cdots, X^{C_{r+1}}_L$ and $U^{C_1},\cdots, U^{C_{r+1}}$, such that $E=E_{XX}\bigcup \nolimits_{i=1}^{r+1} E_i$, in which $E_{i}$ (resp. $E_{XX}$) are the edges between $X_L^{C_i}$ and $U^{C_i}$ (resp. between $X_L$ and $X_R$). Rewrite $X_L\cup X_R\cup U=\{x_1,...,x_{n_V}\}$ and $E=\{e_1,...,e_{n_E}\}$, where $n_V\doteq|X_L\cup X_R\cup U|$ and $n_E\doteq|E|$.  
Define the $(n_V+r)\times (n_E+r)$ augmented incidence matrix $M$ as follows (see Fig. \ref{proof-illustration} for instance): {\small
$$ M_{ij}=\left\{ \begin{aligned}
&= 1, {\text {if}} \ v_i\in \partial(e_j), 1\le i \le n_V, 1\le j \le n_E \\
&= -1, {\text {if}} \ n_V+1\le i \le n_V+r, e_j\in E_{{i-n_v}} \\
&= 1, {\text {if}} \ n_V+1\le i \le n_V+r, j= n_E+i-n_V \\
& =0, {\text {otherwise}},
\end{aligned}\right. $$}where $\partial(e_j)$ represents the vertices in edge $e_j$. Then, matrix $M$ is TU.
\end{proposition}
}

\begin{proposition}\label{unimodular-addcardinality} Consider the bipartite graph $G$ and matrix $M$ in Proposition \ref{unimodular}. Let the $(n_v+r+1)\times (n_E+r)$ matrix $\hat M$ be {\small
$$ \hat M_{ij}=\left\{ \begin{aligned}
&= M_{ij}, {\text {if}} \ 1\le i \le n_V+r, 1\le j \le n_E+r \\
&= 1, {\text {if}} \ i=n_V+r+1, e_j\in \bigcup \nolimits_{l=1}^{r+1} E_l\\
&= -1, {\text {if}} \ i=n_V+r+1, n_E+1\le  j \le n_E+r \\
& =0, {\text {otherwise}}.
\end{aligned}\right. $$}
Then, matrix $\hat M$ is TU.
\end{proposition}

\vspace{-0.5em}
\subsection{LP relaxations and polynomial solvability of ${\cal P}_i|_{i=1}^3$}

\begin{theorem}\label{main-theorem} 
Suppose $(A,B)$ satisfies Assumptions \ref{sc-assump}-\ref{broader-assumption}. Then,  the following LP relaxation ${\cal P}_i^{\rm LP}$ has an integral optimal solution that corresponds to exactly the optimal solution to the ILP ${\cal P}_i^{\rm ILP}$, for $i=1,2$, respectively:
\begin{align}
\min_{y,z} \quad & \sum \nolimits_{(u,v)\in E_{UX}} y_{uv} + |{\cal I}|- \sum \nolimits_{i\in {\cal I}} z_i \tag{${\cal P}^{\rm LP}_1$}\label{LP1} \\
{\rm{s.t.}}\quad & (\ref{C1}), (\ref{C2}), {\rm and}\ (\ref{C3}) \label{CLP1} \\
& 0\le y_{uv} \le 1, \forall (u,v)\in E_{XX}\cup E_{UX} \label{CLP4} \\
& 0 \le z_{i} \le 1, \forall i\in {\cal I}. \label{CLP5}
\end{align}
\begin{align}
\min_{y,z} \quad & \sum \nolimits_{(u,v)\in E_{UX}} w_{uv}y_{uv} + \sum \nolimits_{i\in {\cal I}} (1-z_i)w_i^{\min} \tag{${\cal P}^{\rm LP}_2$}\label{LP2} \\
{\rm{s.t.}}\quad & (\ref{C1}), (\ref{C2}), (\ref{C3}), (\ref{CLP4}), {\rm and}\ (\ref{CLP5}) \label{C6add}
\end{align}
In other words, under Assumptions \ref{sc-assump}-\ref{broader-assumption}, problems ${\cal P}_1$ and ${\cal P}_2$ can be solvable in polynomial time by solving their LP relaxations ${\cal P}_1^{\rm LP}$ and ${\cal P}_1^{\rm LP}$.
\end{theorem}

\begin{proof}
Rewrite the equality (\ref{C1}) as two inequalities:
\begin{equation}\label{equality-inequal} \begin{array}{l}
\sum \nolimits_{(u,v)\in E_{XX}\cup E_{UX}} y_{uv} \le 1, \forall v\in X_L \\
-\sum \nolimits_{(u,v)\in E_{XX}\cup E_{UX}} y_{uv} \le -1, \forall v\in X_L.
\end{array}
\end{equation} Given $(\bar A, \bar B)$ satisfying Assumption \ref{broader-assumption}, let $M$ be defined in Proposition \ref{unimodular} by regarding the vertices of source SCCs $X_i^L|_{i=1}^r$ and the vertex union of non-source SCCs $\bigcup \nolimits_{i=r+1}^{n_c} X_i^L$ as $X_L^{C_i}|_{i=1}^r$ and $X_L^{C_{r+1}}$, respectively. Let $M_{X_L}$ be the rows of $M$ corresponding to $X_L$.
Then, the feasible region of ${\cal P}_i^{\rm LP}$ can be written as $\{x: M_{\rm LP}x\le b_{\rm LP}, x\ge 0\}$, where
$$
{\small M_{\rm LP}=\left[
             \begin{array}{c}
               M \\
               -M_{X_L} \\
               I_{n_E+r} \\
             \end{array}
           \right], b_{\rm LP}=} {\tiny\left[
                                 \begin{array}{c}
                                   1_{n_V\times 1} \\
                                   0_{r\times 1}\\
                                   -1_{n\times 1} \\
                                   1_{(n_E+r)\times 1} \\
                                 \end{array}
                               \right]}.
$$for $i=1,2$.  Since $M$ is TU from Proposition \ref{unimodular}, it follows easily that every square submatrix of $M_{\rm LP}$ still has determinant $0$ or $\pm 1$ (via a similar manner to the proof of Proposition \ref{unimodular}). By definition, $M_{\rm LP}$ is TU, leading to the required assertion.
\end{proof}

\begin{theorem} \label{main-theorem2}
With Assumptions \ref{sc-assump}-\ref{broader-assumption}, the minimum cost $k$-sparsity input selection problem (${\cal P}_3$) can be solved in polynomial time via solving the following LP relaxation ${\cal P}_3^{\rm LP}$.
\begin{align}
\min_{y,z} \quad & \sum \nolimits_{(u,v)\in E_{UX}} w_{uv}y_{uv} + \sum \nolimits_{i\in {\cal I}} (1-z_i)w_i^{\min} \tag{${\cal P}^{\rm LP}_3$}\label{LP3} \\
{\rm{s.t.}} \quad & (\ref{C1}), (\ref{C2}), (\ref{C3}), (\ref{C7}), (\ref{CLP4}), {\rm and}\ (\ref{CLP5}) \ \label{CLP8}
\end{align}
\end{theorem}

\begin{proof}
Similar to the proof of Theorem \ref{main-theorem}, for ${\cal P}_3^{\rm LP}$ in its standard form, the respective constraint matrix is obtained from $\hat M$ after duplicating its rows corresponding to $X_L$ multiplied by $-1$ then adding a unit matrix $I_{n_E+r}$, and thus is TU since $\hat M$ is. The results then follow immediately.
\end{proof}

In light of the theorems above, problems ${\cal P}_i|_{i=1}^3$  can be solved efficiently simply via LP whenever $(A,B)$ satisfies Assumption \ref{broader-assumption}. This means, some off-the-shelf LP solvers could be directly used towards these problems, including the simplex algorithm (although not polynomial time in the worst case, it is quite efficient in practise), the interior point method, the ellipsoid algorithm,  and their state-of-the-art improvements (c.f. \cite{lee2015efficient}) (see footnote 5 for the case when a non-integral optimal solution is returned).  Particularly, it is shown in \cite{vaidya1989speeding} that an LP with $d$ variables can be solved in time $O(d^{2.5}L)$, where $L$ is the number of input bits. For ${\cal P}^{\rm LP}_i$ in its standard form ($i=1,2,3$), upon defining $N\doteq |E_{UX}|+|E_{XX}|$, it has $N+r$ variables, and $L\approx {\log}_2 w_{\max}+{\log}_2 k+1$. Therefore, under Assumption \ref{broader-assumption}, problem ${\cal P}_i$ can be solved in time $O((r+N)^{2.5}L)\to O(N^{2.5}L)$, $i=1,2,3$, noting determining the SCCs and finding $E_{\rm mat}^{\star}$ and $E_{\rm rea}^{\star}$ both have linear complexity in $N+r$.

%

 When restricted to the dedicated input case for problems ${\cal P}_1$ and ${\cal P}_2$, the result above is not a running time improvement compared with the graph-theoretic methods in \cite{S.Pe2016A, A_Olshevsky_2015, pequito2016minimum}. Nevertheless, the power of our LP-based method lies in that it can {\emph{handle more complicated cases}} than the dedicated input one in a unifying manner with polynomial time complexity, remarkably, {\emph{without}} using any combinatorial structure of these problems, except for the SCC decompositions. Therefore, it is also conceptually simple and easy for programmatic implementation. By contrast, it seems unclear how to extend the graph-theoretic methods to problems ${\cal P}_3$, or problems ${\cal P}_1$ and ${\cal P}_2$ beyond the dedicated input case.



 Another significance of Theorems \ref{main-theorem}-\ref{main-theorem2} lies in that, it gives an algebraic, rather than algorithmic proof for the polynomial solvability of the addressed input selection problems under a wide variety of input constraints. So far, it appears that all the identified nontrivial cases that enable polynomial solvability are related to TU. Further, the TU structure allows us to characterize the complexity status of these problems more precisely.

\begin{definition}[Strongly polynomial time]
An algorithm that runs in time (more precisely, the number of elementary arithmetic operations required to exclude this algorithm, including addition, subtraction, multiplication, division and comparison) polynomial of the {\emph{number}} of input items (i.e., dimension of the variables), irrespective of the {\emph{size of the input values}} (namely, the length of the unary representation of the input numerical value), is called a strongly polynomial (time) algorithm.
\end{definition}



\begin{theorem}
With Assumptions \ref{sc-assump}-\ref{broader-assumption}, problems ${\cal P}_1$, ${\cal P}_2$, and ${\cal P}_3$ can be solved in strongly polynomial time.
\end{theorem}

\begin{proof}
From \citep[Theorem 1]{artmann2017strongly}, we know ILPs with constraint matrices that are TU are solvable in strongly polynomial time. The required results follow from this fact and the proven assertions in Theorems \ref{main-theorem} and \ref{main-theorem2} that the constraint matrices of  ${\cal P}^{\rm ILP}_i|_{i=1}^3$ are all TU.
\end{proof}

The theorem above reveals that, with Assumptions \ref{sc-assump}-\ref{broader-assumption}, there exist polynomial algorithms for problems ${\cal P}_1$, ${\cal P}_2$, and ${\cal P}_3$ whose running time depends only on $n,m,N$ and $r$ (for example, the algorithm given in \cite{artmann2017strongly}), but is independent of the size of the sparsity bound $k$ and the input costs $[w_{ij}]$ (thus for a fixed $(A,B)$, the value of $k$ or $[w_{ij}]$ will not affect the running time). This is a stronger conclusion than Theorems \ref{main-theorem}-\ref{main-theorem2}. Note the interior point methods and the ellipsoid algorithms for LPs are usually not strongly polynomial, since their running time might scale with the numerical values of inputs \cite{tardos1986strongly}. 

Finally, when Assumption \ref{broader-assumption} is not satisfied, the corresponding constraint matrices in the ILPs are not necessarily TU. This can be seen from the following example.

\begin{figure}
  \centering
  \includegraphics[width=1.6in]{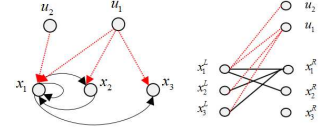}\\
  \caption{System digraph of $(A,B)$ and the associated ${\cal B}(A,B)$ in Example \ref{asp-not-satisfied}. Dotted red edges represent the input links.}\label{plant1}
\end{figure}

\begin{example}\label{asp-not-satisfied} Consider a system $(A,B)$ with its system digraph ${\cal G}(A,B)$ given in Fig. \ref{plant1}, from which we know $r=1$ and $X_1=\{x_1,x_2\}$. This system does not satisfy Assumption \ref{broader-assumption}. For problem ${\cal P}_1$ or ${\cal P}_2$, the corresponding matrix $M$ defined in Proposition \ref{unimodular} reads (not unique s.t. the order of edges)
 {$$\tiny M=\left[
  \begin{array}{ccccccccc}
1&0&0&1&1&0&0&1&0\\
0&1&0&0&0&1&0&0&0\\
0&0&1&0&0&0&1&0&0\\
1&1&1&0&0&0&0&0&0\\
0&0&0&1&0&0&0&0&0\\
0&0&0&0&0&0&0&0&0\\
0&0&0&0&1&1&1&0&0\\
0&0&0&0&0&0&0&1&0\\
0&0&0&0&-1&-1&0&-1&1\\
  \end{array}
\right].$$}It is seen easily that the submatrix of $M$ with rows indexed by $\{1,3,4,7,9\}$ and columns by $\{1,3,6,7,8\}$
has a determinant $-2$. This means $M$ is not TU.
\end{example}


\begin{remark} It is still open whether a general LP admits a strongly polynomial time algorithm \cite{schrijver1998theory}. It is also remarkable that TU is a sufficient condition for an LP to have integral optimal solutions  (in fact, it can be verified the LP relaxation can still solve Example \ref{asp-not-satisfied}).  
 \end{remark}
\vspace{-0.4em}
\subsection{Proof of total unimodularity} \label{proof-TUM}
 The following criterion for TU is needed for our proofs.

\begin{lemma}\label{Ghouila-houri}(Ghouila-Houri Characterization, \cite{lawler2001combinatorial}) A $p\times q$ integral matrix $A=[a_{ij}]$ is TU, if and only if each set $R\subseteq \{1,...,p\}$ can be divided into two disjoint subsets $R_1$ and $R_2$ such that
$$\sum \nolimits_{i\in R_1}a_{ij} -\sum \nolimits_{i\in R_2} a_{ij} \in \{-1,0,1\}, j=1,...,q. $$
\end{lemma}

\begin{figure}
  \centering
  \includegraphics[width=1.4in]{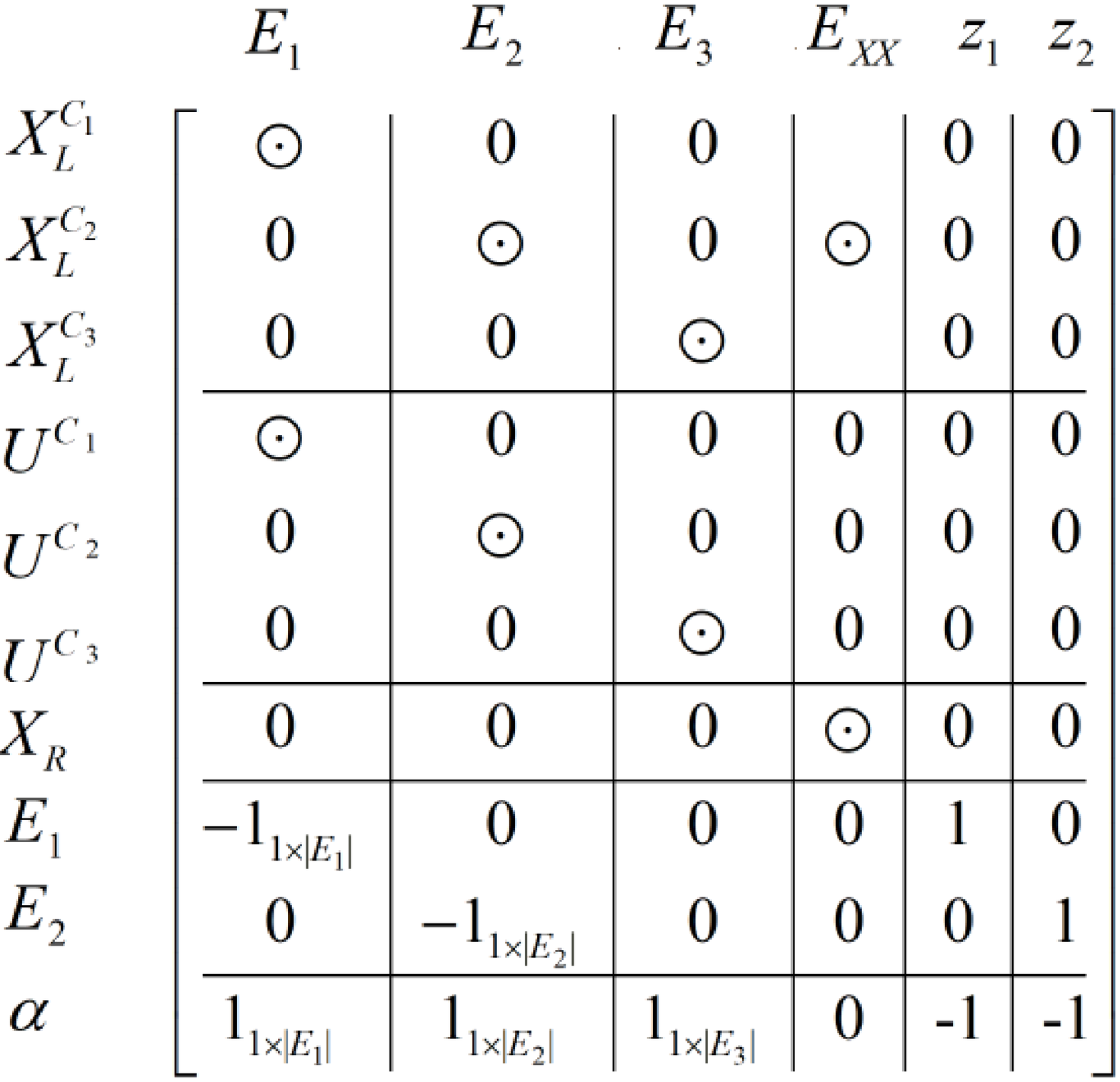}\\
  \caption{Illustration of $M$ and $\hat M$ in the proofs of Propositions \ref{unimodular}-\ref{unimodular-addcardinality} for $r=2$. $\bigodot$ represents a matrix with entries from $\{0,1\}$.}\label{proof-illustration}
\end{figure}

{\bf{Proof of Proposition \ref{unimodular}:}} 
Let $\bar M$ be the matrix consisting of the first $n_E$ columns of $M$. We first prove {\emph{by induction}} that $\bar M$ is TU. Recall that by definition, each column of $\bar M$ indexed by $e\in E_{r+1}\cup E_{XX}$ has exactly $2$ nonzero entries $1$'s, corresponding to the two end vertices of $e$, and each column indexed by $e\in \bigcup \nolimits_{i=1}^r E_i$ has exactly three nonzero entries, among which two $1$'s correspond to the end vertices of $e$ and the third one $-1$ to $E_i$ (see Fig. \ref{proof-illustration} for illustration).  For the beginning of the induction, the claim is certainly true for any $1\times 1$ submatrix of $\bar M$. Assume the claim holds true for all $(k-1)\times (k-1)$ submatrices of $\bar M$. Let $M'$ be a $k\times k$ submatrix of $\bar M$. If there is a column of $M'$ which contains none nonzero entry, then certainly $\det M'=0$. If there is a column of $M'$ which contains exactly one nonzero entry, then $\det M'= \pm \det M''$, where $M''$ is obtained from $M'$ after deleting the respective row and column containing the aforementioned entry. Hence, $\det M'\in \{0,\pm 1\}$ by induction.

Otherwise, each column of $M'$ contains at least $2$ nonzero entries. We consider two cases. In the first case, $M'$ does not contain $-1$'s, which means each column contains exactly two $1$'s. As $G$ is bipartite, the rows of $M'$ can be partitioned into two sets $R_1$ and $R_2$, such that for each of its columns, there is exactly one $1$ in each set. Then, for each column $j\in \{1,...,k\}$ of $M'$, $\sum \nolimits_{i\in R_1}M'_{ij}-\sum \nolimits_{i\in R_2}M'_{ij}=0$, meaning $\det M'=~0$.


In the second case, $M'$ has some rows which contain $-1$'s.  We consider two subcases. {\bf Subcase i)}: Each column of $M'$ contains $-1$'s. Since each column of $M'$ contains at most two $1$'s and at least one $1$, as well as exactly one $-1$, summing up all the elements in each column will get $1$ or~$0$.

{\linespread{1.4} \selectfont
{\bf Subcase ii)}: $M'$ contains some columns which do not have $-1$'s. Suppose the columns of $M'$ that contain $-1$'s are indexed by $E_{s1}$, and the columns not containing $-1$'s but corresponding to the subset of $\bigcup \nolimits_{i=1}^r E_i$ are indexed by $E_{s2}$, and the remaining columns are indexed by $E_{s3}$ ({we may alternatively use the corresponding edge or vertex to denote the respective column or row of a submatrix}).  Without losing any generality, assume that the $E_{s1}$ is a subset of $\bigcup \nolimits_{i=1}^{r_1} E_i$, and $E_{s2}$ is a subset of $\bigcup \nolimits_{i=r_1+1}^r E_i$, $1\le r_1 \le r$. Note $G$ is bipartite, and for each $i\in \{1,...,r+1\}$, the rows of $\bar M$ corresponding to $U^{C_i}$ do not have $1$'s except in its columns indexed by $E_i$. Consequently, we can partition the rows of $M'$ into $R_1,R_2,...,R_r$, $R_L$, $R_R$, and $R_{-1}$, so that $R_i$ contains all rows of $M'$ indexed by $U^{C_i}$, $i=1,...,r$ (some of $R_i$'s may be empty), $R_L$ contains all rows of $M'$ indexed by $X_L$, $R_R$ contains all rows of $M'$ indexed by $U^{C_{r+1}}\cup X_R$, and all rows containing $-1$'s are in $R_{-1}$.
  Observe that, for each column of $M'$ that are not indexed by $\bigcup \nolimits_{i=1}^r E_i$, there is exactly one $1$ in $R_L$ and in $R_R$ (recalling each column of $M'$ contains at least two nonzero entries). Additionally, for each column of $M'$ indexed by $\bigcup \nolimits_{i=1}^r E_i$, there are at most two $1$'s and at least one $1$ in $\bigcup \nolimits_{i=1}^r R_i \cup R_L$, as well as at most one $-1$ in $R_{-1}$. Therefore, for each column $j\in E_{s1}$ of $M'$, it holds
{\scriptsize
$$\underbrace{\sum \limits_{i\in R_L}M'_{ij}}_{0,1}+\underbrace{\sum \limits_{k=1}^{r_1} \sum \limits_{i\in R_k}M'_{ij} }_{0,1}\!-\!\underbrace{\sum \limits_{k=r_1+1}^{r} \sum \limits_{i\in R_k}M'_{ij} }_{0}\!-\!\underbrace{\sum \limits_{i\in R_R}M'_{ij}}_{0}\!+\! \underbrace{\sum \limits_{i\in R_{-1}}M'_{ij}}_{-1}=0,1,$$}
and for each column $j\in E_{s2}$ of $M'$
{\scriptsize
$$\underbrace{\sum \limits_{i\in R_L}M'_{ij}}_{1}\!+\!\underbrace{\sum \limits_{k=1}^{r_1} \sum \limits_{i\in R_k}M'_{ij} }_{0}\!-\!\underbrace{\sum \limits_{k=r_1+1}^{r} \sum \limits_{i\in R_k}M'_{ij} }_{1}\!-\!\underbrace{\sum \limits_{i\in R_R}M'_{ij}}_{0}\!+\! \underbrace{\sum \limits_{i\in R_{-1}}M'_{ij}}_{0}=0,$$}
and for each column $j\in E_{s3}$ of $M'$,
{\scriptsize
$$\underbrace{\sum \limits_{i\in R_L}M'_{ij}}_{1}\!+\!\underbrace{\sum \limits_{k=1}^{r_1} \sum \limits_{i\in R_k}M'_{ij} }_{0}\!-\!\underbrace{\sum \limits_{k=r_1+1}^{r} \sum \limits_{i\in R_k}M'_{ij} }_{0}\!-\!\underbrace{\sum \limits_{i\in R_R}M'_{ij}}_{1}\!+\! \underbrace{\sum \limits_{i\in R_{-1}}M'_{ij}}_{0}=0.$$}By Lemma \ref{Ghouila-houri}, in all these subcases, we get $\det M'\in \{0,\pm 1\}$. This proves that $\bar M$ is TU. }

Note $M$ is obtained after adding the matrix $I_{r}$ next to the lower right corner of $\bar M$. Any square submatrix $M'$ of $M$ that contains some nonzero entries of $I_{r}$ has a structure as follows
{\footnotesize$$ M'=\left[
\begin{array}{c|c}
  M''  &    0     \\
    \hline
       \ddots   & I_{r'}
\end{array}
\right] \ {\rm or} \ \left[
\begin{array}{c|c}
  \ddots  &    I_{r'}     \\
    \hline
       M''   & 0
\end{array}
\right]\ ,$$}where $M''$ is obtained by deleting the rows and columns belonging to $I_r$, $r'\le r$. Therefore, $\det M'=\pm \det M''\in \{0,\pm 1\}$ by the  TU of $\bar M$. By definition, $M$ is TU. \hfill $\square$

  {\bf{Proof of Proposition \ref{unimodular-addcardinality}:}}  As $\hat M$ is obtained by adding a row to $M$, and $M$ is TU, it suffices to show that every square submatrix $\hat M'$ of $\hat M$ that contains nonzero entries from the last row of $\hat M$ (denoting this row by $\alpha$; see Fig. \ref{proof-illustration}) is TU. First, consider the case where $\hat M'$ does not contain any nonzero elements in the last $r$ rows of $M$ (denoting these rows by $\beta_{1},\cdots, \beta_{r}$ from the top down, respectively). Notice $\alpha$ has $1$'s in its columns indexed by $\bigcup \nolimits_{i=1}^{r+1}E_i$ and $-1$ in its $n_E+1$ to $n_E+r$ columns, and $0$'s elsewhere. Also observe that for each row of $M$ in the set indexed by $\bigcup \nolimits_{i=1}^{r+1}U^{C_i}$, there are no nonzero entries except in the columns indexed by $\bigcup \nolimits_{i=1}^{r+1}E_i$. This means, by changing the sign of the last row of $\hat M'$, we can find an assignment of signs for each row of $\hat M'$ so that the sum of their signed rows equals a row vector with entries in $\{0,\pm 1\}$, in a way similar to that for the square submatrix $M'$ of $M$ in the proof of Proposition~\ref{unimodular}.

  Now, consider the case where $\hat M'$ contains nonzero elements from $\alpha$ and some $\beta_i$'s simultaneously. Without sacrificing any generality, suppose $\hat M'$ contains nonzero elements from $\beta_1,...,\beta_{r_1}$, $1\le r_1 \le r$.  Observe that $(\alpha + \sum \nolimits_{i=1}^{r_1}\beta_i)_j = 1$ when $j$ is indexed by $\bigcup \nolimits_{i=r_1+1}^{r+1} E_i$,  $(\alpha + \sum \nolimits_{i=1}^{r_1}\beta_i)_j= -1$ when $j=n_E+r_1+1,...,n_E+r$, and $(\alpha + \sum \nolimits_{i=1}^{r_1}\beta_i)_j = 0$ elsewhere. Introduce a $(n_V+1)\times (n_E+r)$ matrix $\tilde M$, consisting of the first $n_V$ rows of $M$ and its last row being $\alpha+ \sum \nolimits_{i=1}^{r_1}\beta_i$.  Notice that for each row of $M$ in the set indexed by $\bigcup \nolimits_{i=r_1+1}^{r+1} U^{C_i}$, there are no nonzero elements except in the rows indexed by $\bigcup \nolimits_{i=r_1+1}^{r+1} E_i$. Following the similar reasoning to the proof of Proposition \ref{unimodular}, it turns out every submatrix of $\tilde M$ is TU. Consequently, there is an sign assignment for each row of $\hat M'$, such that the sum of their signed rows yields a row vector with entries in $\{0,\pm 1\}$ (in which the rows containing nonzero elements from $\alpha$ and $\beta_1,...,\beta_{r_1}$ always have the same sign). By Lemma \ref{Ghouila-houri}, this means $\hat M$ is TU. \hfill $\square$


%
%
%



\section{Illustrative example} \label{example}
\begin{figure}
  \centering
  \includegraphics[width=1.7in]{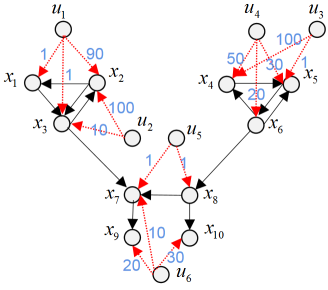}\\
  \caption{System digraph of $(A,B)$. Dotted red edges represent the input links, with the numbers in green being their costs.}\label{plant2}
\end{figure}

This section provides an example to illustrate the LP methods reported in this paper.\footnote{The code of this example is available at \url{https://github.com/
Yuanzhang2014}.}  Consider a system $(A,B)$ with its system digraph ${\cal G}(A,B)$ given in Fig. \ref{plant2}. The costs of available input links are given therein, too.

It is easy to see this system satisfies Assumptions \ref{sc-assump}-\ref{broader-assumption}. This system consists of $6$ SCCs, among which $X_1=\{x_1,x_2,x_3\}$ and $X_2=
\{x_4,x_5,x_6\}$ are the vertex sets of two source SCCs (thus $|{\cal I}|=r=2$). Moreover, $w_1^{\min}=w_2^{\min}=1$.  Introducing variables $\{y_{e}: e\in E_{UX}\cup E_{XX}\}$ and $\{z_1,z_2\}$, we can formulate Problems ${\cal P}_1$, ${\cal P}_2$, and ${\cal P}_3$ as the LPs ${\cal P}_1^{\rm LP}$, ${\cal P}_2^{\rm LP}$, and ${\cal P}_3^{\rm LP}$, respectively. Particularly, for problem ${\cal P}_3$, we set $k=3$.

With the help of the Matlab LP solver {\emph{linprog}}, for the LP ${\cal P}_1^{\rm LP}$, it is found the optimal solution is $y_{e}=1$ for $e=(u_1,x_2^L),(u_4,x_4^L)$, $(x_2^R,x_1^L), (x_1^R,x_3^L)$, $(x_3^R,x_7^L),(x_7^R,x_9^L)$, ~\\ $(x_4^R,x_5^L), (x_5^R,x_6^L), (x_6^R,x_8^L)$, and $(x_8^R,x_{10}^L)$, and $y_e=0$ otherwise; in addition, $z_1=z_2=1$. This means the optimum to problem ${\cal P}_1$ is $2$, and the corresponding optimal input selection is $\{(u_1,x_2),(u_4,x_4)\}$ (the total cost is $140$, which is the minimum cost that can be achieved with $2$ input links).

For the LP ${\cal P}_2^{\rm LP}$, it turns out the optimal solution corresponds to $E^{\star}_{\rm mat}=\{(u_1,x_1^L),(u_3,x_5^L), (u_5,x_8^L), (u_6,x_7^L)\}$ and $E^{\star}_{\rm rea}=\emptyset$ ($E^{\star}_{\rm mat}$ and $E^{\star}_{\rm rea}$ are defined in Theorem \ref{solution}), with the optimum $13$. This indicates the optimum to problem ${\cal P}_2$ is $13$, and the corresponding optimal input selection is $\{(u_1,x_1),(u_3,x_5), (u_5,x_8), (u_6,x_7)\}$.

Similarly, for problem ${\cal P}_3$ with $k=3$, it is attained that the LP ${\cal P}_3^{\rm LP}$ achieves its optimum $52$ with $E^{\star}_{\rm mat}=\{(u_1,x_1^L),(u_4,x_4^L), (u_5,x_7^L)\}$ and $E^{\star}_{\rm rea}=\emptyset$. This means, the optimum to problem ${\cal P}_3$ is $52$, and the corresponding optimal solution is $\{(u_1,x_1),(u_4,x_4), (u_5,x_7)\}$.

The above three solutions also indicate that there might be a trade-off between the sparsity and the cost of the optimal input selections for structural controllability, which highlights the significance of problem ${\cal P}_3$. Say, with a smaller sparsity bound, the cost of the corresponding optimal input selection tends to be bigger.

\section{Conclusions} \label{conclusion}
This paper investigates three related cost-sparsity induced optimal input selection problems for structural controllability in the non-dedicated constrained input case. We first formulate these problems as equivalent ILPs, and then show under the said source-SCC grouped input constraint, those ILPs could be solved efficiently by their LP relaxations using the off-the-shelf LP solvers. We further show those problems are strongly polynomially solvable. We do this by proving that the corresponding constraint matrices of the ILPs are TU. In this way, we provide an alternatively algebraic approach, conceptually different
from the graph-theoretic ones, for these problems under the addressed condition, which, contains all the existing known polynomially solvable (nontrivial) ones as special cases. It is expected that some graph-theoretic algorithms for these problems might be attained from the primal-dual algorithms for the corresponding LPs (c.f. \citep[Chap 4]{lawler2001combinatorial}), which could be the future work.




%
\bibliographystyle{elsarticle-num}
{\footnotesize
\bibliography{yuanz3}
}
\end{document}